\documentclass{revtex4}  
\usepackage[latin1]{inputenc}
\usepackage{amsmath}
\usepackage{amsfonts}
\usepackage{amssymb}
\usepackage{graphicx,color}

\begin{document}

\title{Quantum holography with biphotons of high dimensionality}

\author{Fabrice Devaux$^{1}$, Alexis Mosset$^1$, Florent Bassignot$^2$ and Eric Lantz$^1$}
\affiliation{$^1$ Institut FEMTO-ST, D\'epartement d'Optique P. M. Duffieux, UMR 6174 CNRS \\ Universit\'e Bourgogne Franche-Comt\'e, 15b Avenue des Montboucons, 25030 Besan\c{c}on - France\\
$^2$ FEMTO Engineering, 15b Avenue des Montboucons, 25030 Besan\c{c}on - France}

\date{\today}
%

\begin{abstract}
We report results of two-photon quantum holography where spatial information stored in phase holograms is retrieved by measuring quantum spatial correlations between two images formed by spatially entangled twin photons with a dimensionality of 1790 in the two-dimensional transverse space. In our experiments, the entire flux of spontaneous down converted photons illuminates the phase holograms and the photons of pairs signal-idler transmitted by the holograms are detected separately in far field on two electron-multiplying charge coupled device cameras.       
\end{abstract}


\maketitle

High-dimensionality spatial entanglement allows access to large Hilbert spaces, with applications in numerous fields of quantum optics, like quantum lithography \cite{boto_quantum_2000}, quantum computation \cite{tasca_continuous-variable_2011} or quantum ghost imaging \cite{pittman_optical_1995}. By itself, a source of quantum light issued from spontaneous down conversion (SPDC) appears as incoherent, preventing the formation of an image of the spatial spectrum of an object (a transparency) in the Fourier plane. However, coincidence imaging of the pairs of twin photons allows this spatial spectrum to be retrieved, as demonstrated in experiments like demonstration of spatial antibunching \cite{nogueira_experimental_2001}, observation of two-photon speckle patterns \cite{peeters_observation_2010}, or transfer of the angular spectrum of the transparency modulating the pump beam \cite{monken_transfer_1998}. All these experiments measured coincidences between two single-photon counting modules scanned on the signal and idler images. These procedures are time-consuming, even if improved by compressive-sensing  \cite{howland_efficient_2013}, and use a very little part of the incident photons, leading to potential loopholes  \cite{lantz_optimizing_2014} if applied to the demonstration of basics properties of entanglement like the Einstein-Podolsky-Rosen (EPR) paradox \cite{howell_realization_2004}. 

Because of these drawbacks, imaging with single-photon sensitive cameras has became more and more popular and allows massively parallel coincidence counting. Examples of experiments include sub-shot noise imaging \cite{jedrkiewicz_detection_2004, brida_experimental_2010} using a low noise CCD camera, demonstration of a high degree of EPR paradox \cite{moreau_einstein-podolsky-rosen_2014} and of transmission of biphotons through a non unitary object\cite{reichert_biphoton_2017}  using electron-multiplying CCD cameras (EMCCD), unity contrast EPR-based ghost imaging with an intensified CCD camera \cite{aspden_epr-based_2013}, as well as holography of a single photon with an intensified CMOS camera \cite{chrapkiewicz_hologram_2016}.

In this paper, we report coincidence imaging of bidimensional phase holograms using two EMCCD cameras. As in ref. \cite{nogueira_experimental_2001,peeters_observation_2010}, no single photon image is formed in the Fourier plane, while the cross correlation of the images allows a coherent image to be retrieved in the far field, with an equivalent wavelength equal to half the signal or idler wavelength, as in \cite{santos_resolution_2003}.

For a sufficiently thin crystal, it can be assumed that the two photons of a pair are created at the same random place. This assumption is equivalent to neglecting the uncertainty in the image plane due to phase matching conditions. Hence, the two-photon spectral wave function of SPDC emitted from a thin crystal pumped by a monochromatic beam of angular frequency $\omega_p$ and of amplitude $E_p(\mathbf{r})$ is given by \cite{saleh_duality_2000}:
\begin{equation}\label{eq1}
\psi (\mathbf{r}_1,\mathbf{r}_2;\omega_s) \propto \int E_p(\mathbf{r}) h_s(\mathbf{r_1},\mathbf{r};\omega_s)h_i(\mathbf{r_2},\mathbf{r};\omega_p-\omega_s)d\mathbf{r},
\end{equation}
where $\mathbf{r_1}$ and $\mathbf{r_2}$ are transverse positions in the plane of separate detectors ($EMCCD_1$ for the signal and $EMCCD_2$ for the idler) and $\mathbf{r}$ is a coordinate in the image plane of the crystal where the hologram lies. $h_s(\mathbf{r}_1,\mathbf{r};\omega_s)$ and $h_i(\mathbf{r}_2,\mathbf{r};\omega_p-\omega_s)$ are the impulse response functions of the separate linear imaging systems for the signal and the idler beams, respectively.
 
\begin{figure}[h!]
\centering
\includegraphics[width=12cm]{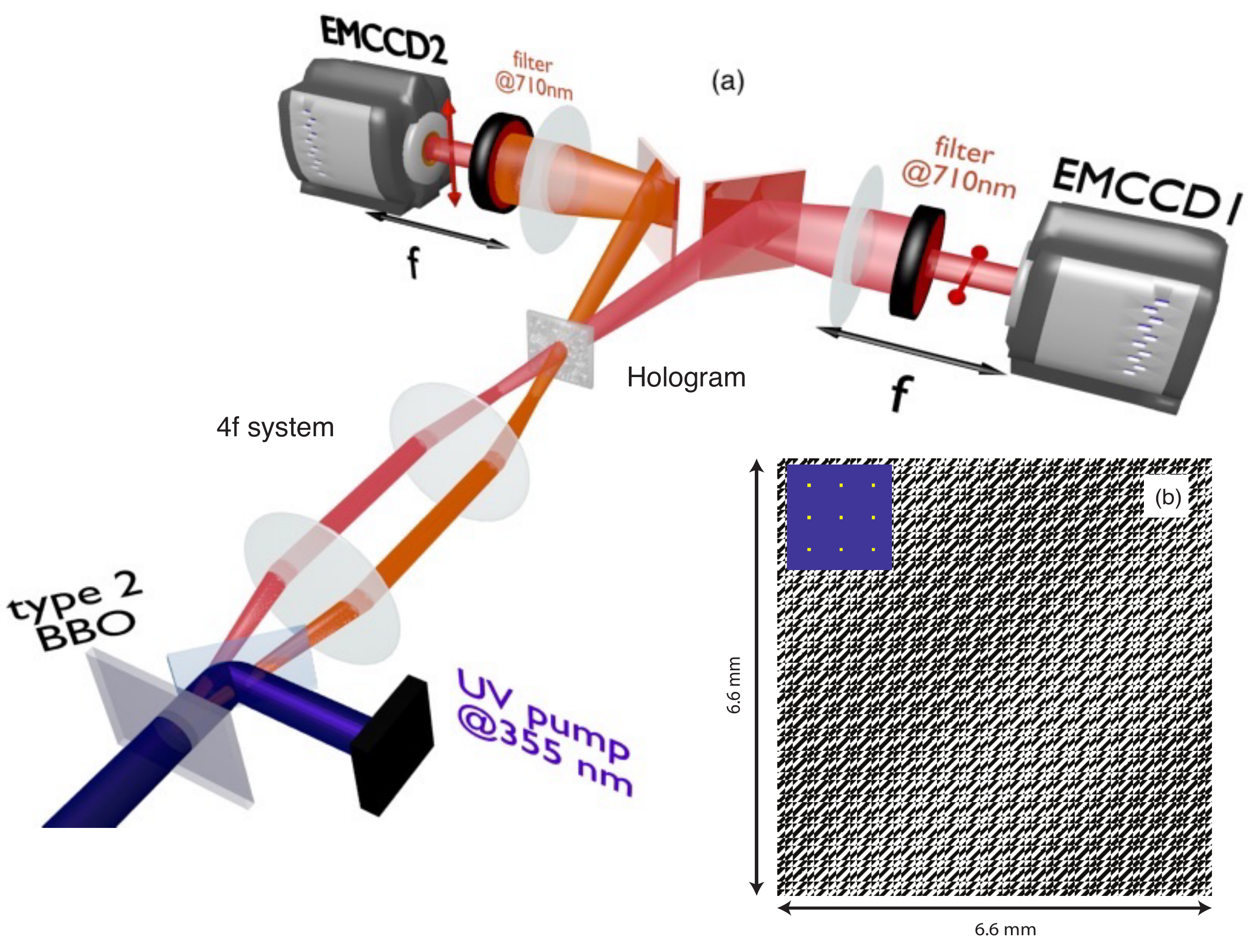}
\caption{(a) Experimental setup : Twin photon pairs at $710\,nm$ are generated by SPDC in a type-II BBO crystal. The crystal is imaged with a $4-f$ optical system on a binary phase hologram engraved on a glass slide. The photons signal and ilder transmitted by the hologram are then naturally separated by free space propagation thanks to the walk-off. They are then detected and resolved spatially in the far field on two EMCCD cameras used in photon counting mode . (b) Pattern of the $\left(0-\frac{\pi}{2}\right)$ binary phase hologram. The insert represents the pattern encoded in the hologram: an array of 9 Dirac peaks.}
\label{setup}
\end{figure}

Now, let us name $t(\mathbf{r})=e^{i\varphi(\mathbf{r})}$ the transmission of the phase hologram with a phase modulation $\varphi(\mathbf{r})$ and let us make some assumptions. First, we assume that the hologram is thin and planar. Secondly, the binary phase hologram is designed in such a way that the $\pm 1$ diffraction orders are centered, in a far field, on $\pm 6\,mm^{-1}$ spatial frequencies which are much smaller than the $64\,mm^{-1}$ phase-matching bandwidth (FWHM) of the type-II BBO crystal (see the paragraph dedicated to the experimental set-up), in agreement with the above assumption of neglecting the effect of imperfect phase-matching. Moreover, because SPDC is detected in a narrow band around degeneracy, the biphoton state is assumed to be monochromatic ($\omega_i=\omega_p-\omega_s=\omega_s$). Thirdly, in our experimental setup (Fig. \ref{setup}a), as the hologram is placed in the near field of the crystal and because all photons are collected by the $4-f$ imaging system, in Eq. \ref{eq1}, we can consider that the separate optical systems are formed only by the hologram and the two identical Fourier transform optical systems ($2-f$ systems). Then, the impulse responses are given by: 
\begin{equation}\label{eq2}
h_{s,i} \left( \mathbf{r_{1,2}}, \mathbf{r};\omega\right) = t_{s,i}(\mathbf{r})\frac{e^{-2ikf}}{i\lambda f}e^{-\frac{ik}{f}\mathbf{r_{1,2}} \mathbf{r}},
\end{equation}
where $k$ and $\lambda$ are the signal or idler wave number and wavelength (at the degeneracy $k_s=k_i=k$, $\lambda_s=\lambda_i=\lambda$).
Then,  Eq. \ref{eq1} becomes:
\begin{equation}\label{eq3}
\psi (\mathbf{r_1},\mathbf{r_2}) \propto \int E_p(\mathbf{r})t_s\left( \mathbf{r}\right)t_i\left( \mathbf{r}\right) e^{-\frac{ik}{f}(\mathbf{r_1}+\mathbf{r_2})\mathbf{r}} d\mathbf{r}=\int E_p(\mathbf{r})t^2\left( \mathbf{r}\right)e^{-\frac{ik}{f}(\mathbf{r_1}+\mathbf{r_2})\mathbf{r}} d\mathbf{r}
\end{equation}
Hence, the experimental two-photon coincidence rate at two positions in the separate detection planes is given by:
\begin{equation}\label{eq4}
G^{(2)}\left( \mathbf{r_1},\mathbf{r_2}\right) =\left|\psi (\mathbf{r_1},\mathbf{r_2})\right|^2 \propto\left|\widetilde{E_p}\left( \frac{\mathbf{r_1}+\mathbf{r_2}}{\lambda f}\right) \ast \widetilde{t^2}\left( \frac{\mathbf{\mathbf{r_1}+\mathbf{r_2}}}{\lambda f}\right)\right| ^2.
\end{equation}
where $\ast$ denotes the convolution product and $\,\widetilde{}\,$ the bidimensional Fourier-transform operator. In this expression, the transmission of the hologram is squared, unlike in classical coherent imaging. In consequence, while binary phase holograms are usually designed with a $\left(0-\pi\right) $ phase step for efficient restitution with coherent light, a $\left(0-\frac{\pi}{2}\right)$ phase step must be engraved when a biphoton source is used or, equivalently, the $\left(0-\pi\right) $ phase step must be engraved by considering a halved wavelength \cite{santos_resolution_2003}. 

The experimental set-up is illustrated in Fig. \ref{setup}a. Photon pairs are generated via SPDC in a type-II geometry in a $0.8\,mm$ long $\beta$-barium borate ($\beta$-BBO) crystal pumped at $355\,nm$. The pump pulses are provided by a passively Q-switched Nd:YAG laser (330 $ps$ pulse duration, 27 $mW$ mean power, 1 $kHz$ repetition rate and 1.6 $mm$ FWHM beam diameter). The crystal (i.e. near field of twin photons) is imaged with a $4-f$ imaging system on a binary phase hologram with a transversal and an angular magnification of -1 and the entire flux of spontaneous down converted light illuminates the hologram. Fig. \ref{setup}b shows the binary pattern engraved on a glass slide to create the phase hologram and the insert corresponds to the pattern encoded in the hologram : an array of 9 Dirac peaks. The  phase holograms are designed to produce off-axis patterns and the binary hologram, i.e. diffractive optical elements (DOE) \cite{oshea_diffractive_2003}, gives a restitution of the orignal pattern at $\pm 1$ diffraction orders. The engraving depth of the holograms is adjusted to produce a $\left(0-\frac{\pi}{2}\right)$ binary phase modulation at $710\,nm$ in order to optimize the diffraction efficiency of the hologram with the biphotons source. 
The cross-polarized signal and ilder beams transmitted by the hologram are then naturally separated by free space propagation thanks to the walk-off. Lastly, photons of pairs are detected and resolved spatially in the far-field on two EMCCD cameras (Andor iXon3) used in photon counting mode \cite{lantz_multi-imaging_2008}. Before detection, photons pairs emitted around the degeneracy are selected by narrow-band interference filters centered at $710\,nm$ ($\Delta\lambda\backsimeq 4\,nm$). The vertical and horizontal red arrows symbolize the polarization directions of the signal and idler beams. 

The protocol for measuring spatial momentum correlations between twin images is the same as the one we proposed for measuring EPR paradox \cite{moreau_einstein-podolsky-rosen_2014,lantz_einstein-podolsky-rosen_2015} and temporal ghost imaging  \cite{denis_temporal_2017} with twin images. For a set of twin images, we apply a thresholding procedure to convert the gray scales into binary values that correspond to 0 or 1 photon. Then, a spatial coincidence correlation function is obtained by calculating the normalized cross-correlation of photodetection images, after subtraction from these images of their deterministic part (i.e. the mean of the images).      

\begin{figure} [ht!]
\centering
\includegraphics[width=12cm]{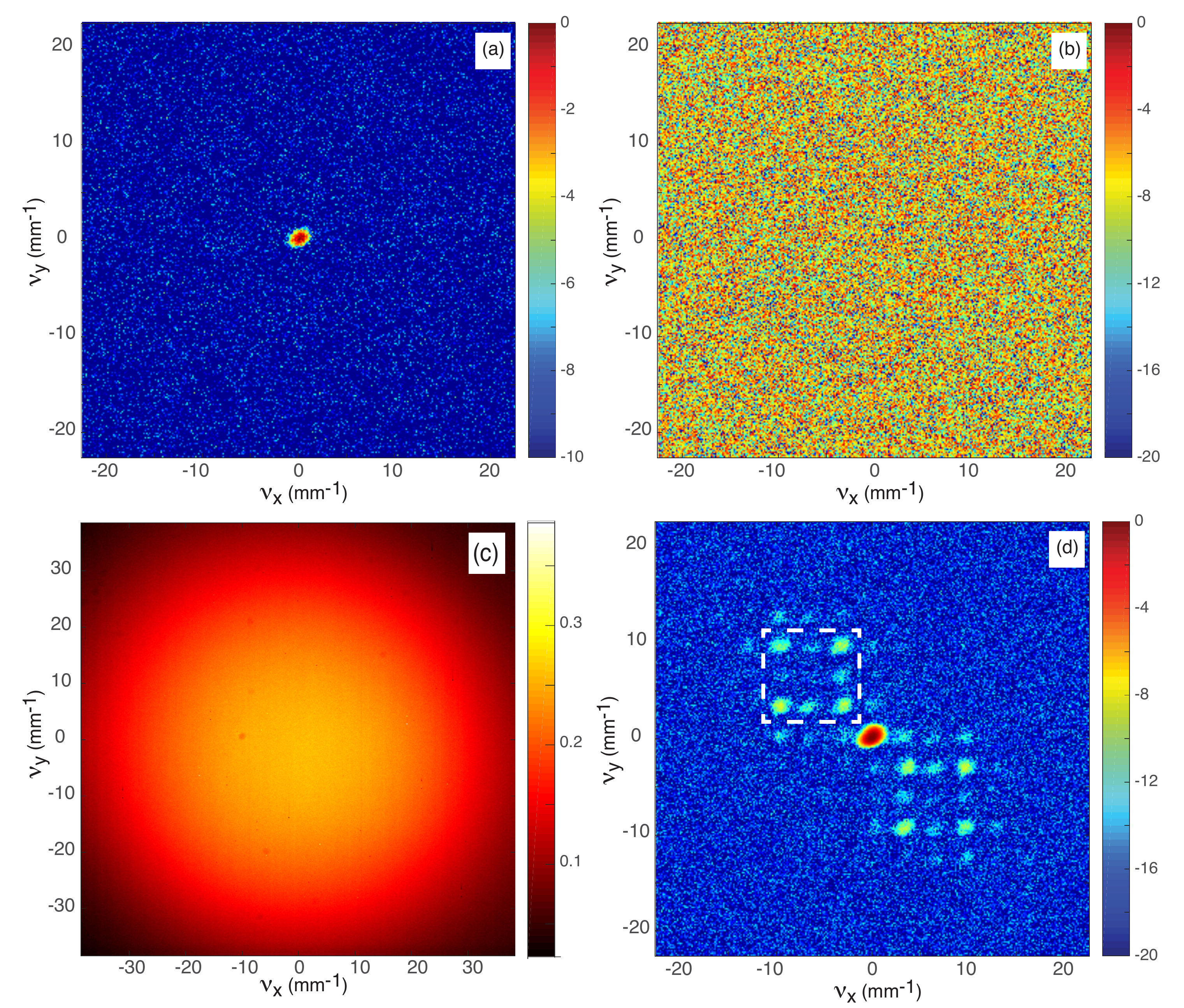}
\caption{Without the hologram : in $dB$, normalized cross-correlation in momentum between 100 twin images (a) and images that do not share pump pulses (b). With the hologram: average photon number in single far-field images (signal or idler) of SPDC (c) and restored hologram formed by the normalized cross-corrrelation in momentum, given in $dB$, calculated over 80000 twin images (d). The white dotted squares indicate the location of the original pattern encoded in the hologram.}
\label{fig2}
\end{figure}

First, the hologram is removed from the experimental setup. From the normalized cross-correlation in momentum calculated with a set of 100 twin images (Fig. \ref{fig2}a), we measure the width of the correlation peak, expressed in standard deviations, which gives, in spatial frequency units: $\sigma_{\nu_x}=0.69\,mm^{-1},\,\sigma_{\nu_y}=0.59\,mm^{-1}$. With $\sigma_\phi=27\,mm^{-1}$ the standard deviation deduced from the $64\,mm^{-1}$ FWHM of the phase matching function (Fig. \ref{fig2}c), we can  estimate roughly the whole dimensionality $V$, i.e. the Schmidt number of the bi-photon wave function in the two-dimensional transverse space :

\begin{equation}\label{eq5}
V=\frac{\sigma_\phi^2}{\sigma_{\nu_x}\sigma_{\nu_y}}\approx 1790
\end{equation}
This value has to be compared with the theoretical value given by \cite{devaux_towards_2012}:
\begin{equation}\label{eq6}
V_{th}=\left( \frac{2\pi0.69}{1.89}\right)^2\frac{\sigma^2_{pump}}{\lambda_sL\left(n_s^{-1}+n_i^{-1}\right)}\approx 3500,
\end{equation}
where $L$ is the crystal thickness, $\sigma_{pump}\approx 0.68\,mm$ is the standard deviation deduced from the FWHM of the gaussian pump beam and $n_s$, $n_i$ the reffractive indices of the BBO crystal. These rough calculations give values of the same order of magnitude and confirm the high dimensionality of the biphoton wavefunction. The discrepancy between these values is probably due to the telescope's geometric aberrations which are at the origin of a widening of the correlation peak and thus to a reduction of the effective Schmidt number. We also calculate the integral of the normalized correlation peak, i.e. the degree of correlation, that is equal to 0.25. This value represents the ratio between the number of photons detected in pairs and the total number of photons. This result is also consistent with the equivalent quantum efficiency of the entire detection system which includes the quantum efficiency of the cameras and the transmission coefficients of the various optical components (filters, lenses, dichroic mirrors) \cite{lantz_einstein-podolsky-rosen_2015}. Finally, we verified that there is no deterministic correlations between images that do not share pump pulses (Fig. \ref{fig2}b). 

We now put the phase hologram back into the experimental set-up. In Fig. \ref{fig2}c showing the far-field mean spatial distribution of photons signal (or idler) transmitted by the hologram, we can observe that the spatial information encoded in the hologram is not retrieved, because of the incoherent nature of SPDC \cite{peeters_observation_2010,monken_transfer_1998}. In contrast and in good agreement with Eq. \ref{eq4}, when cross-corrrelation in momentum between twin images is calculated, the spatial distribution of two-photon coincidence rate exhibits a pattern (Fig. \ref{fig2}d), where appears at the $\pm 1$ diffraction orders the original pattern encoded in the hologram: an array of 9 Dirac peaks (in white dotted squares). Some additional periodically distributed peaks are also visible due to the binary character of the hologram. In order to improve the signal-to-noise ratio (SNR) of the retrieved pattern, this cross-corrrelation image is calculated over 80000 twin images. Although the engraving depth of the hologram is adjusted with an accuracy of about $10\%$ to give a $\left(0-\frac{\pi}{2}\right)$ phase step, the correlation peak corresponding the  0-order diffraction of the hologram is much more intense than the $\pm1$ diffraction orders, unlike in the perfect coherent image. From the integral of the whole normalized correlation pattern, the degree of correlation of twin images is estimated to 0.20. It is smaller than the degree of correlation measured without the hologram because the transverse momenta of some photon pairs transmitted by the hologram are greater than the maximum sampling spatial frequency imposed by the sensor dimensions. The results presented here correspond to the best position of the hologram, minimizing the level of the correlation peak at the 0-order diffraction. 
  \begin{figure} [ht!]
  	\centering
  	\includegraphics[width=12cm]{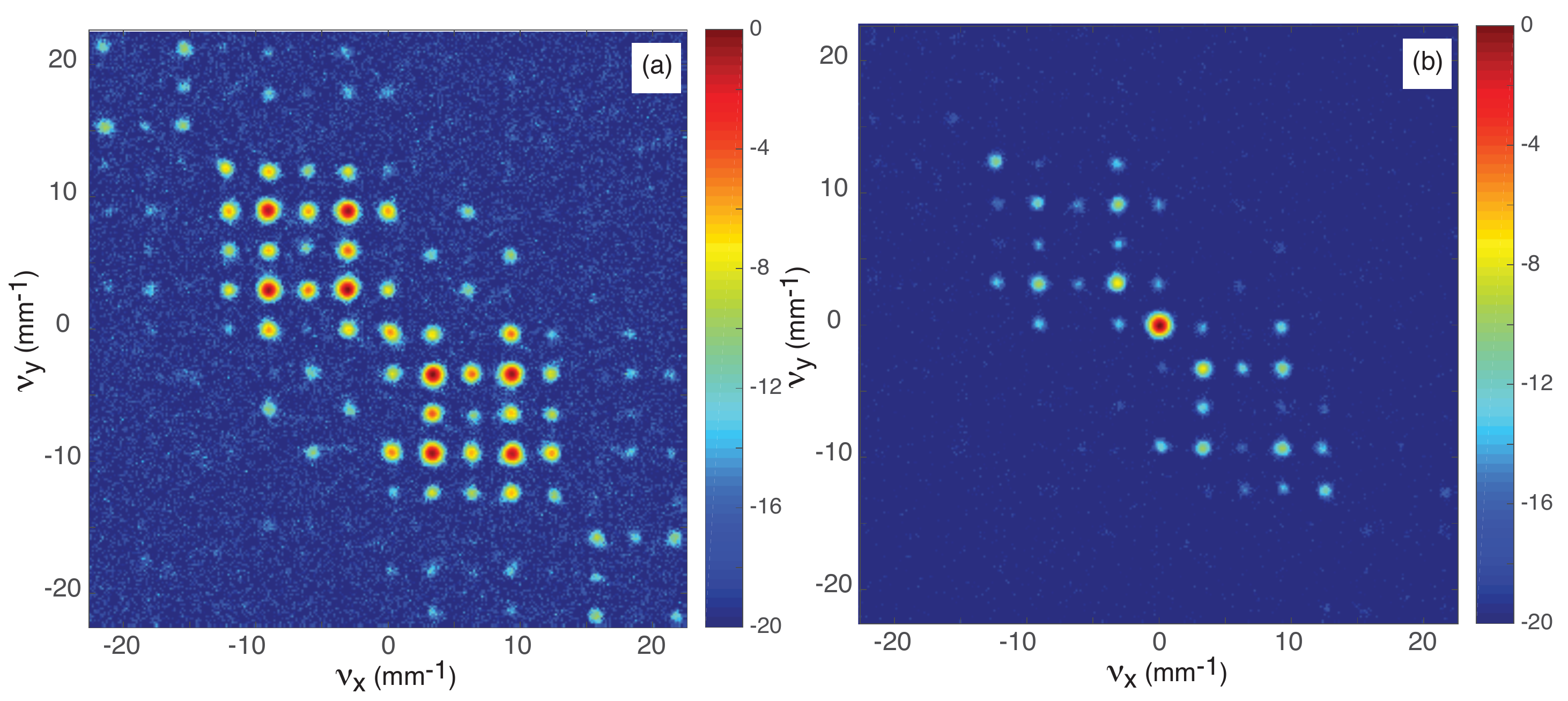}
  	\caption{ Normalized averaged cross-corrrelation of images issued from stochastic simulations,  sampled and scaled for direct comparison with the experimental results of Fig. \ref{fig2}d. No defocusing (a), $1.5\, mm$ defocusing (b).}
  	\label{fig3}
  \end{figure}

To explain the large amplitude of the 0-order correlation peak, we have assumed a slight defocusing ($1.5\, mm$) of the image plane of the crystal with respect to the position of the hologram. This defocusing could be due in part to the geometric aberrations of the $4-f$ imaging system. Eq. \ref{eq1} remains valid but the impulse-responses are no more given by Eq. \ref{eq2}. A formalism involving for each beam two impulse-responses (one from the crystal to the object and one from the object to the image) could be developed \cite{abouraddy_entangled-photon_2002}, leading to double integrals which must be calculated for each couple of pixels $(\mathbf{r_1},\mathbf{r_2})$. For a bidimensional image, the computation time would scale at the eighth power of the number of pixels in one dimension, which is prohibitively long. Fortunately, stochastic simulations based on the Wigner formalism \cite{lantz_spatial_2004} allow here accurate results when repeated several thousand times and averaged. Fig. \ref{fig3}a shows the cross-correlation image for no defocusing, which appears quite different of the experimental image. On the other hand, introducing  $1.5\, mm$ of defocusing leads to a simulated image very close of the experimental one. Note that these simulations also take into account the propagation in the crystal, resulting, for the mean one-photon image, in a phase-matching cone in good agreement with Fig. \ref{fig2}c.

  \begin{figure} [ht!]
  	\centering
  	\includegraphics[width=12cm]{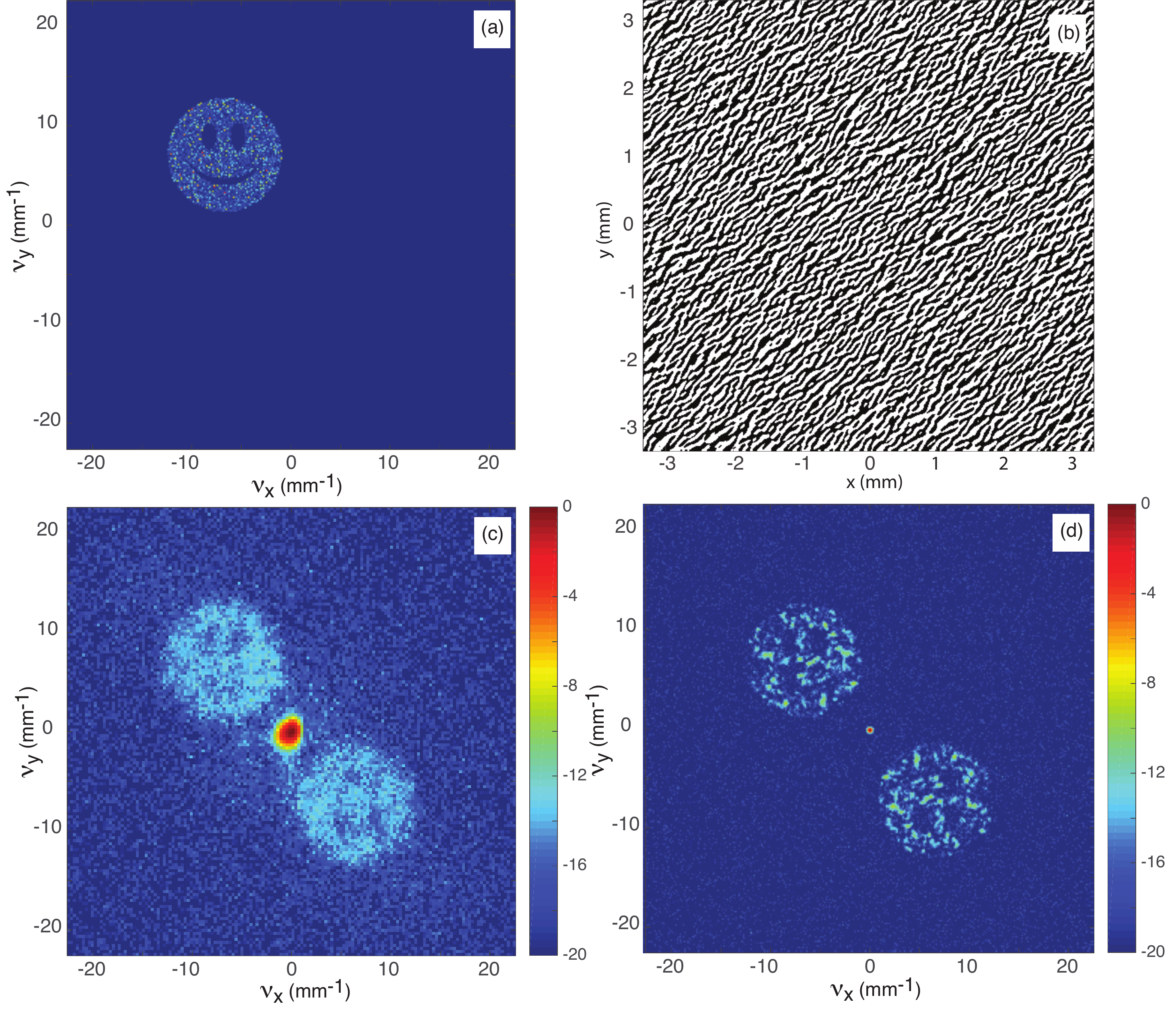}
  	\caption{ (a) Picture encoded in the DOE : a $10\, mm^{-1}$ diameter "Smiley face" modulated by a deterministic speckle pattern. (b) Binary pattern of the DOE. (c) Restored hologram formed by the normalized cross-corrrelation calculated over 270000 twin images. (d) Normalized averaged cross-correlation of images issued from 10000 stochastic simulations where no defocusing is considered.}
  	\label{fig4}
  \end{figure}

Finally, we used an another DOE (Fig. \ref{fig4}b) designed to produce a "smiley face" of $10\, mm^{-1}$ diameter modulated by a deterministic speckle pattern (Fig. \ref{fig4}a). Figures \ref{fig4}c and \ref{fig4}d show for comparison the restored hologram formed by the normalized cross-corrrelation calculated over 270000 twin images and the normalized averaged cross-correlation of images issued from 10000 stochastic simulations where no defocusing is considered, respectively. Although spatial coincidences reproduce the original pattern, we can observe that the resolution of the smiley face is strongly limited by the size of the speckle grains that compose it. Indeed, according to Eq. \ref{eq4}, these grains are the result of the convolution between the speckle grains of the initial pattern and the intercorrelation peak observed at the 0-order diffraction, of width proportional to the inverse of the width of the pump beam in the near field. From the integral of the whole normalized correlation pattern, the degree of correlation of twin images is estimated to 0.25. Because coincidences between twin photons are spread over large areas, it is necessary to cumulate a much larger number of realizations in order that the pattern formed by the coincidences emerges from the background noise. For the same reason as with the previous hologram, we can observe that the 0-order diffraction peak concentrates a significative part of the spatial coincidences between the twin images.   
  
\section{Conclusion}

To summarize, we have shown that two photon imaging potentially allows coherent manipulation  of light in complex situations like holography. These results generalize previous demonstrations where the biphoton image was a one-dimensional interference pattern created by a double-slit \cite{nogueira_experimental_2001} or a  one dimensional speckle scattered by a rough surface \cite{peeters_observation_2010}. Unlike these previous experiments, all the light is used, preventing loopholes due to the selection of a small part of the photons and allowing full bi-dimensional manipulation of high dimensionality biphoton states, with potential applications in present hot topics, like boson sampling.
  
\section*{Funding}
This work was partly supported by the French "Investissements d'Avenir" program, project ISITE-BFC (contract ANR-15-IDEX-03) and the RENATECH network and its FEMTO-ST MIMENTO technological facility.


\end{document}